\documentclass[a4paper,11pt]{article}
\usepackage{pos}
\usepackage{adjustbox}
\usepackage{amsmath}
\usepackage{graphicx}
\usepackage{booktabs}
\usepackage{subfigure}

\title{Gluonic Energy Momentum Tensor Form Factors of the Proton}

\author*[a]{Zein-Eddine Meziani}

\affiliation[a]{Physics Division, Argonne National Laboratory,\\
9700 South Cass Avenue, Lemont, USA}

\emailAdd{zmeziani@anl.gov}

\abstract{Gravitational form factors (GFFs), defined through the matrix elements of the energy-momentum tensor, provide critical insights into the internal structure of nucleons and nuclei. In particular, their Fourier transforms in the Breit frame yield spatial distributions of mass, pressure, and shear force densities associated with both quark and gluon constituents.
This work presents recent measurements of near-threshold $J/\psi$ photoproduction on the proton, performed in Hall C at Jefferson Lab, utilizing data from the electronic decay channels of the $J/\psi$. These results enable the extraction of gluonic gravitational form factors (gGFFs), offering a novel probe of the gluon dynamics within the nucleon.
The analysis employs a holographic QCD framework to interpret the threshold behavior of the cross sections and to facilitate the extraction of the gGFFs. The implications of these measurements are discussed in the context of upcoming experimental programs, including the near-threshold electro- and photoproduction studies with SoLID at Jefferson Lab and the $\Upsilon$ production program at the Electron-Ion Collider using the ePIC detector. These future efforts are expected to significantly improve the precision of gGFF determinations and provide essential tests of their universality across different kinematic regimes. }

\FullConference{The XVIth Quark Confinement and the Hadron Spectrum Conference (QCHSC24)\\
 19-24 August, 2024\\
 Cairns Convention Centre, Cairns, Queensland, Australia\\}

\begin{document}
\maketitle

\section{Introduction}
The study of nucleon and nuclear structure has entered a transformative phase, driven by new experimental and theoretical developments that allow access to gravitational form factors (GFFs). These fundamental quantities encode intrinsic mechanical properties of hadrons, including mass, pressure, and shear force distributions. When extended to nuclei, GFFs offer promising new avenues for exploring the dynamics of nuclear matter.
The Fourier transform of specific combinations of GFFs in the Breit frame reveals the spatial distribution of energy, analogous to how electromagnetic form factors determine the charge distribution within the proton. A detailed discussion of the proton's GFFs, particularly from Deep Virtual Compton Scattering (DVCS), and the extraction of quark-sector contributions can be found in Ref.~\cite{Burkert:2023wzr}.
However, gluons, the mediators of the strong force and a dominant source of the nucleon's mass, play an equally essential role. In this presentation, I will focus on the gravitational form factors associated with gluons.

The matrix element of the QCD energy-momentum tensor for quarks or gluons reads~\cite{Ji:1996ek}:
\begin{align}
&\langle p_f, s_f | T_{q,g}^{\mu,\nu}(0) | p_i,s_i \rangle = \\  \nonumber
&\bar u(p_f,s_f) \displaystyle{\Bigl ( } A_{q,g}(t) \gamma^{ \{ \mu} P^{\nu \} }  + B_{q,g} \frac{iP^{ \{ \mu }\sigma^{\nu \} \rho}\Delta_{\rho}}{2M_N} 
+ C_{g,q} \frac{\Delta^{\mu}\Delta^{\nu} -g^{\mu\nu}\Delta^2}{M_N} + \bar C_{q,g}(t) M_N g^{\mu,\nu} \displaystyle{ \Bigr )} u(p_i,s_i)
\end{align}

\noindent where \((p_i , s_i )\) and \((p_f, s_f )\) correspond to the momentum and polarization of the incoming
and outgoing nucleon, respectively, \( P= (p_i + p_f )/2,~t= \Delta^2=(p_f-p_i)^2 \). \( A_{q,g}(t) \), \(B_{q,g}(t)\), \(C_{q,g}(t) \), and \( \bar C_{q,g}(t) \) are the quarks and gluons GFFs of the nucleon. It is worth noting that \( J(t)= 1/2 [ A(t)+B(t) ] \) where \(J_{q+g}(0)=1/2\) is the total spin of the proton. 

The first experimental extraction of the quark gravitational form factor (GFF) \( D(t) = 4C(t) \) was achieved through analyses of Deep Virtual Compton Scattering (DVCS) data, leading to the determination of the quark pressure distribution within the proton in the Breit frame~\cite{Burkert:2018bqq}. To gain further insight into the internal dynamics of the proton, particularly the distribution of quark energy contribution to the proton mass, it is essential to extract the quark GFF \( A_q(t) \). This, in turn, requires a precise determination of the generalized parton distribution (GPD) \( H_q \) over a wide kinematic range in the four-momentum transfer \( Q^2\), the Bjorken scaling variable \(x\), and the momentum transfer \(t \), at fixed skewness \( \xi \) = 0, using the DVCS process. Ongoing experimental efforts at Jefferson Lab~\cite{Arrington:2021alx}, and future measurements at the Electron-Ion Collider (EIC)~\cite{AbdulKhalek:2021gbh}, aim to fulfill this objective and advance our understanding of the quark sector's contribution to the nucleon's gravitational structure.

Several frameworks exist for decomposing the proton mass; a comprehensive review can be found in Ref.\cite{Lorce:2021xku} and references therein. Among them, Ji's mass decomposition\cite{Ji:1994av}, inspired by the virial theorem, provides a physically intuitive partitioning of the nucleon mass into quark and gluon components. Within this framework, the total gluonic contribution is further separated into two distinct parts: a tensor glueball component, carrying the quantum numbers of the graviton \( 2^{++} \), and a scalar glueball component, associated with dilaton-like quantum numbers \(0^{++} \). The scalar component is particularly important, as it is linked to the nucleon matrix element of the trace anomaly of the QCD energy-momentum tensor and is scale invariant. This term encapsulates the breaking of classical scale invariance in quantum chromodynamics and defines the characteristic energy scale of the strong interaction.

In this presentation, I will focus on the determination of the gluons GFFs using the measurement of near threshold photoproduction of the \( J/\psi \) meson in Hall C experiment \( J/\psi\text{-}007 \) at Jefferson Lab and will provide the gluonic mass distribution profile in the Breit frame as well as the gluon pressure profile, both extracted from the experimental data.

\section{The \( \boldsymbol{J/\psi}\text{-}007  \) experiment at Jefferson Lab}
The Jefferson Lab experiment E12-16-007~\cite{Meziani:2016lhg} also known as \(J/\psi \text{-}007\) proposed to search for the LHCb pentaquark, was performed in Hall C at Jefferson Lab and consisted of using the CEBAF electron beam at an energy of 10.6 GeV passing through an 8.5\% copper radiator upstream of a central pivot of rotation of two high momentum spectrometers (HMS \& SHMS). The mixed beam of electrons and photons  passed through a 15 cm liquid hydrogen target located at this pivot producing $J/\psi$ particles. The $J/\psi$ e$^+$e$^-$ and $\mu^+ \mu^-$ pair decays were detected in the HMS and SHMS, respectively. While the recoil proton was not measured estimates of the incoherent background, where an extra pion is produced, as well as the Bethe-Heitler contamination were found to be negligible and consistent with our description of the invariant mass of the leptons pair spectrum identifying the $J/\psi$ peak. After detectors efficiency corrections for each spectrometer, and combined spectrometers acceptance corrections at different kinematics settings, the doubly-differential cross sections where unfolded in two-dimensional bins of different  photon energy $E_{\gamma}$ and four momentum transfer to the proton $t$. The description of the experiment and resulting differential cross sections are shown in ref.~\cite{Duran:2022xag} in comparison with several cross section models~\cite{Kharzeev:2021qkd,Mamo:2019mka,Guo:2021ibg,Hatta:2018ina,Hatta:2019lxo,Sun:2021gmi} .

\section{Gluonic gravitational form factors \( \boldsymbol{A_g} \) and \(\boldsymbol{C_g} \) determination}

To extract the gluonic form factors \( A_g(t) \) and \( C_g(t) \), we performed two-dimensional fits across photon energy \( E_{\gamma} \) and momentum transfer \( t \) of the differential cross sections using the holographic QCD model developed by Mamo and Zahed (M-Z)~\cite{Mamo:2019mka,Mamo:2022eui}. A next-to-leading order generalized parton distribution (GPD) analysis of data from the $J/\psi$-007 and GlueX experiments~\cite{GlueX:2023pev}, employing Bayesian inference techniques to extract these gluonic form factors, is also detailed in Ref.~\cite{Guo:2025jiz}. The emphasis on the holographic approach stems from its inherently non-perturbative nature, making it particularly suitable for modeling near-threshold $J/\psi$ photoproduction.

In the M-Z holographic model the cross section is expressed as follows~\cite{Mamo:2019mka,Mamo:2022eui}:
\begin{equation}
\frac{d\sigma}{dt} =  \frac{{\cal N}^2 e^2}{64\pi(s-m_N^2)^2}\times \frac{[ A_g(t) + \eta^2 D_g(t) ]^2}{A_g^2(0)}\times F(s,t, M_{J/\psi},m_N)
\times \frac{(-2t+8m_N^2)}{4m_N^2} ,
\end{equation}

\noindent $A_g(t)$ and $D_g(t)$ shapes are fully calculated in this model (see \cite{Mamo:2022eui}). The normalization constant  ${\cal N}^2e^2$ = (7.768)$^2$ nb/GeV$^{6}$ is taken from ~\cite{Mamo:2019mka}. For a consistent comparison with the lattice QCD calculations of Ref.~\cite{Pefkou:2021fni} we chose a tripole functional form for both $A_g(t)$ and $C_g(t)$ form factors 
\begin{eqnarray}
A_g(t) = A_g(0)\left ( 1- \frac{t}{m_A^2} \right )^{-3},~~~~~~~~~C_g(t) = C_g(0)\left ( 1- \frac{t}{m_C^2} \right )^{-3}
\end{eqnarray}
in our fitting procedure of the differential cross sections, where $A_g(0)$ is the average momentum fraction carried by the gluons in the nucleon and determined by the CT18 global parametrization~\cite{Hou:2019efy} of the world DIS data, $A_g(0)$ = $\langle x \rangle_g =0.414\pm 0.008$. The three other parameters, namely $m_A$, $C(0)$ and $m_C$ are free parameters determined by the two-dimensional fit of the differential cross section data of the $J/\psi -007$ experiment.

\begin{figure}[t]
\includegraphics[width=\textwidth]{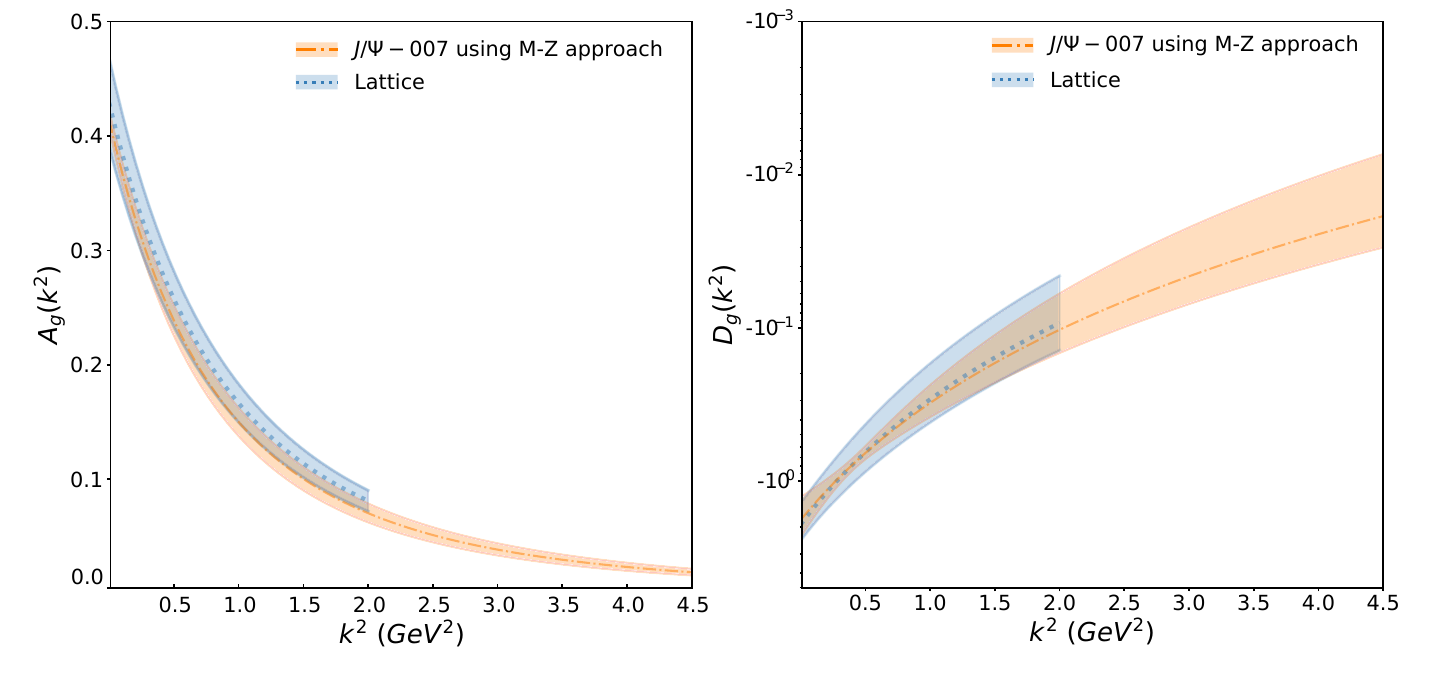}
\caption{
Left panel: The $A_g(k^2)$ form factor ($k^2=|t|$) extracted from our two-dimensional cross section data in the holographic QCD approach~\cite{Mamo:2019mka,Mamo:2021krl} (orange dash-dot curve), compared to the latest lattice calculation~\cite{Pefkou:2021fni} (blue dotted curve). In all cases the form factors used are of tripole functional form. The shaded areas show the corresponding uncertainty bands. Right panel: The extracted $D_g(k^2)=4C_g(k^2)$ form factor with the same color scheme as the left panel.}
\label{fig:gff-ac}
\end{figure}

The mass radius is expressed as
\begin{equation}
    \langle r_{m}^2 \rangle_g  =  \frac{1}{A_g (0)}  6 \left . \frac{dA_g (t)}{dt} \right \vert_{t=0} - 6  \frac{C_g (0)}{M^2_N}~~  =~~ \frac{18}{m_A^2} - 6\frac{1}{M^2_N} \frac{C_g (0)}{A_g (0)}
\label{eqn:m_radius}
\end{equation}

\begin{table*}[ht!]
\caption{ The gluonic GFFs fit parameters and proton mass radius 
determined from the $J/\psi-007$ experiment~\cite{Prasad:2024} through a two-dimensional fit using the holographic QCD approach~\cite{Mamo:2019mka,Mamo:2021krl,Mamo:2022eui}. The corresponding proton mass radius is also reported according to eq.~(\ref{eqn:m_radius}).
In all cases we used the tripole functional functional form  for all GFFs. We compare these results to the latest lattice calculations~\cite{Pefkou:2021fni}.}
\begin{adjustbox}{width=\linewidth}
\begin{tabular}{cccccc}
\toprule
Theoretical approach & $\chi^2$/n.d.f &$m_A$ (GeV) &  $m_C$ (GeV) & $C_g(0)$ &$\sqrt{\langle r_m^2\rangle}_g$ (fm) \\
\midrule
Holographic QCD  & 0.925 &1.575$\pm$0.059 & 1.12$\pm$0.21 & -0.45$\pm$0.132 & 0.755$\pm$0.067 \\
\midrule
Lattice & & 1.641$\pm$ 0.043 & 1.07$\pm$ 0.12 & -0.483$\pm$ 0.133 & 0.7464$\pm$0.055  \\
\bottomrule
\end{tabular}
\end{adjustbox}
\label{jpsi:fitparams-one}
\end{table*}

We note that the holographic QCD approach gives results consistent with the lattice QCD calculations results~\cite{Pefkou:2021fni} . 

To perform our fit and define the radius listed in Table~\ref{jpsi:fitparams-one} we have neglected $B_g(t)$ and ignored $\bar C_g(t)$. The holographic QCD approach is a non-perturbative approach that seems to agree well with lattice QCD, although the lattice calculations are still being performed away from the pion physical mass (m$_{\pi}$=400 MeV).  Measurements of $J/\psi$ electro and photo-production on a proton have been proposed and approved at Jefferson Lab using the SoLID~\cite{JeffersonLabSoLID:2022iod} while measurements of both electro and photo-production using a smaller dipole size, namely $\Upsilon$, are part of the science program at the EIC~\cite{AbdulKhalek:2021gbh,Gryniuk:2020mlh}. Both of these proposed  future measurements would enhance our confidence on the extraction of the GFFs with higher statistical precision and minimal theoretical uncertainty due to controlled approximations.

\section{Gluon mechanical properties in the Breit frame}

Of course, there is more information in the gluons form factors  compared to just the gluon mass radius. Therefore, to provide the full gluon energy density profile and other mechanical properties, we perform a Fourier transform of combinations of the corresponding form factors and provide mass, pressure and shear forces densities as a function of distance $r$ in the Breit frame. Here, again we have no other  choice but to assume $B_g(t)\sim 0$ and ignore $\bar C_g(t)$ for now.  

\begin{figure}[hb!]
\includegraphics[width=\textwidth]{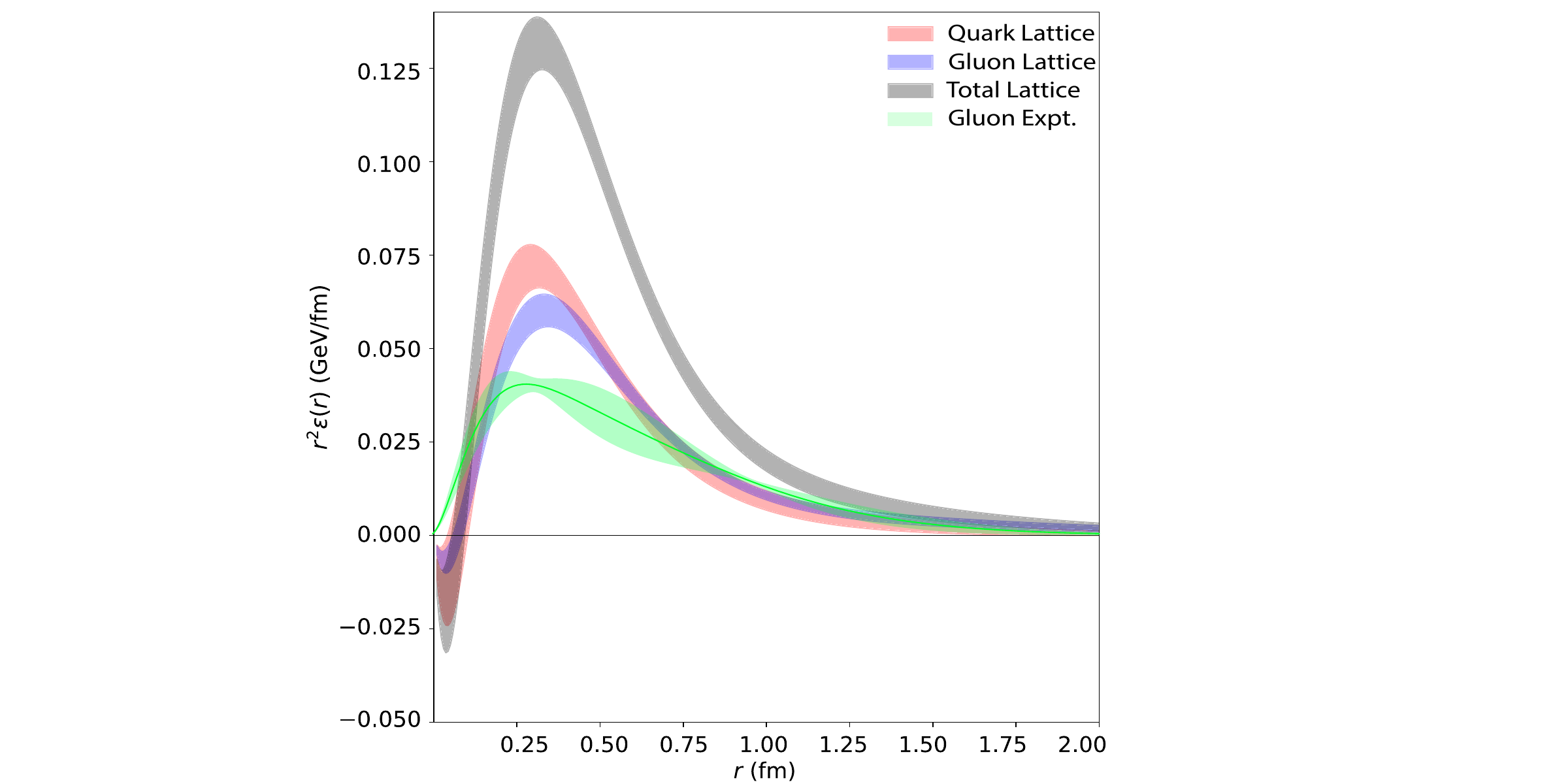}
\caption{The gluon energy density $ r^2 \epsilon$ contribution in the Breit frame,  according to ~\cite{Lorce:2018egm}, with \( A \) and \( D \) GFFs extracted using the holographic QCD approach~\cite{Mamo:2019mka,Mamo:2021krl,Mamo:2022eui} (Green curve). The other bands are the more recent lattice results from~\cite{Hackett:2023rif}. The shaded area shows the corresponding uncertainty band for every contribution. Note that the lattice result uses a dipole-dipole combination of the form factors}
\label{fig:ft-epsi}
\end{figure}

For the gluons mass density profile we used the expression~\cite{Lorce:2018egm}:
\begin{eqnarray}
\epsilon_g(r) &=& M \left \{ A^{FT}_g(r) + \bar{C}^{FT}_g(r) + \frac{1}{4M^2}\frac{1}{r^2}\frac{d}{dr} \left (r^2  \frac{d}{dr} \left [ B^{FT}_g(r)- D^{FT}_g(r)\right ] \right ) \right \}  \nonumber \\
&=& M \left \{ A^{FT}_g(r) - \frac{1}{4M^2}\frac{1}{r}\frac{d}{dr} \left ( r^2 \frac{d}{dr} D^{FT}_g(r) \right ) \right \}
\end{eqnarray}
where \( D^{FT}_g (r)\) and \( D^{FT}_g (r) \) are Fourier transform of \( A(t) \) and \(D(t) \) given by 

\begin{eqnarray}
    A^{FT}_g (r) &=& \int \frac{d^3 \boldsymbol{\Delta}}{(2\pi)^3} e^{-i \boldsymbol{\Delta} \cdot \boldsymbol{r}} A_g(\Delta^2) = A_g(0) \frac{m_A^3}{32\pi}(1+m_A r) e^{-m_A r} \\
    D^{FT}_g (r) &=& \int \frac{d^3 \boldsymbol{\Delta}}{(2\pi)^3} e^{-i \boldsymbol{\Delta} \cdot \boldsymbol{r}} D_g(\Delta^2) = D_g(0) \frac{m_C^3}{32\pi}(1+m_C r) e^{-m_C r}
\end{eqnarray}

Figure~\ref{fig:ft-epsi} shows the profile mass density compared to the latest lattice result~\cite{Hackett:2023rif}. Here $A_g (r)$ and $D_g (r)$ are determined using experiment $J/\psi-007$ data while setting $B_g (r) = 0$ while ignoring $\bar C_g (r)$ consistent with lattice QCD calculations. However, it is worth noting that its contribution subtract strength to $A_q$ while adding strength to $A_g$ since  $\bar C_q (r)$ is negative according to Ref.~\cite{Hatta:2018sqd,Tanaka:2022wzy} and $\bar C_g (r) = -\bar C_q (r)$. Using \( \bar C_g \) will only make the gluon energy contribution larger.

\begin{figure}[hb!]
\includegraphics[width=\textwidth]{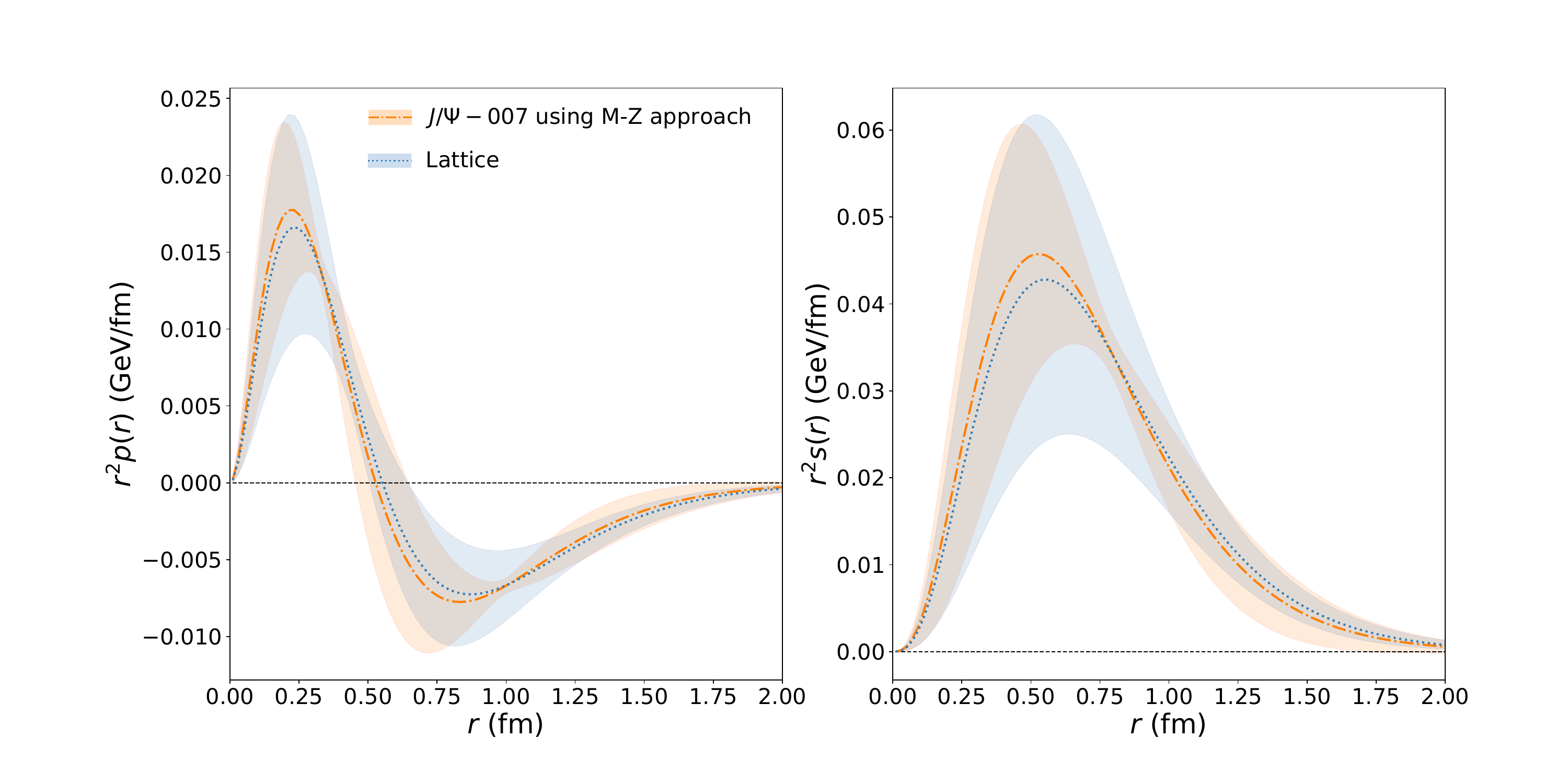}
\caption{Left panel: The $r^2p_g(r)$ pressure in the Breit frame in the holographic QCD approach~\cite{Mamo:2019mka,Mamo:2021krl,Mamo:2022eui} (orange dash-dot curve), compared to the lattice calculation~\cite{Pefkou:2021fni} (blue dotted curve). In this lattice calculation the functional form of the \( D(t) \) form factor is a tripole is a tripole. The shaded areas show the corresponding uncertainty bands. Right panel: The extracted shear forces density $r^2s_g(r)$ of the gluons in the same Breit frame and color scheme as the left panel.}
\label{fig:ft-ps}
\end{figure}

 The pressure $r^2p(r)$ and the shear forces $r^2s(r)$ densities in the Breit frame involve only the \( D \) form factur and are given by~\cite{Lorce:2018egm}:
\begin{equation}
    r^2p(r)=\frac{1}{6M} \frac{d}{dr} \left ( \frac{d}{dr}D^{FT}(r) \right ),~~~
    r^2s(r)= -\frac{1}{4M} r^3 \frac{d}{dr}\left ( \frac{1}{r} \frac{d}{dr}D^{FT}(r) \right )
\end{equation}
Here we used the tripole functional form for $C_g(t)$ form factor to compare directly to the lattice calculations of Ref.~\cite{Pefkou:2021fni}. The experimental gluons results~\cite{Prasad:2024} of pressure and shear forces densities in the Breit frame as a function of distance $r$ are shown in Fig.~\ref{fig:ft-ps}.
  
Finally, it is worth mentioning that  the statistical precision of these results will be  dramatically improved using the SoLID~\cite{JeffersonLabSoLID:2022iod} detector in Hall A at Jefferson Lab. Furthermore, the universal aspect of the extracted gluons GFFs in the threshold region will ultimately be confirmed or refuted at the EIC using  $J/\psi$ electroproduction at large center of mass energies and wide range of $t$. and through the electro- and photo-production of $\Upsilon$ near-threshold~\cite{AbdulKhalek:2021gbh,Gryniuk:2020mlh}.

\section{Acknowledgment}
I thank the organizers for the opportunity to present this work. This work is supported in part by the department of Energy Office of Science, Office of Nuclear Physics under contract DE-AC02-06CH11357.


\begin{thebibliography}{99}



\bibitem{Burkert:2023wzr}
V.~D.~Burkert, L.~Elouadrhiri, F.~X.~Girod, C.~Lorc\'e, P.~Schweitzer and P.~E.~Shanahan,
\emph{Colloquium: Gravitational form factors of the proton,}
Rev. Mod. Phys. \textbf{95} (2023) no.4, 041002
doi:10.1103/RevModPhys.95.041002, [{\tt hep-ph/2303.08347}].

\bibitem{Ji:1996ek}
X.~D.~Ji,
\emph{Gauge-Invariant Decomposition of Nucleon Spin,}
Phys. Rev. Lett. \textbf{78}, 610-613 (1997) doi:10.1103/PhysRevLett.78.610
[{\tt hep-ph/9603249}].

\bibitem{Burkert:2018bqq}
V.~D.~Burkert, L.~Elouadrhiri and F.~X.~Girod,
\emph{The pressure distribution inside the proton,'}
Nature \textbf{557} (2018) no.7705, 396-399, doi:10.1038/s41586-018-0060-z

\bibitem{Arrington:2021alx}
J.~Arrington, 
\textit{et al.},\emph{Physics with CEBAF at 12 GeV and future opportunities,} Prog. Part. Nucl. Phys. \textbf{127}, 103985 (2022)
doi:10.1016/j.ppnp.2022.103985, [{\tt nucl-ex/2112.00060}].

\bibitem{AbdulKhalek:2021gbh}
R.~Abdul Khalek, \textit{et al.},
\emph{Science Requirements and Detector Concepts for the Electron-Ion Collider: EIC Yellow Report,}
Nucl. Phys. A \textbf{1026} (2022), 122447 doi:10.1016/j.nuclphysa.2022.122447, [{\tt physics.ins-det/2103.05419}].

\bibitem{Lorce:2021xku}
C.~Lorc\'e, A.~Metz, B.~Pasquini and S.~Rodini,
\emph{Energy-momentum tensor in QCD: nucleon mass decomposition and mechanical equilibrium,}
JHEP \textbf{11}, 121 (2021), doi:10.1007/JHEP11(2021)121, [{\tt hep-ph/2109.11785}].

\bibitem{Ji:1994av}
X.~D.~Ji,
\emph{A QCD analysis of the mass structure of the nucleon,}
Phys. Rev. Lett. \textbf{74}, 1071-1074 (1995)
doi:10.1103/PhysRevLett.74.1071 [{\tt hep-ph/9410274}].

\bibitem{Meziani:2016lhg}
Z.-E.~Meziani, S.~Joosten, M.~Paolone, E.~Chudakov, M.~Jones, K.~Adhikari, K.~Aniol, W.~Armstrong, J.~Arrington and A.~Asaturyan, \textit{et al.}
\emph{A Search for the LHCb Charmed 'Pentaquark' using Photo-Production of $J/{\psi}$ at Threshold in Hall C at Jefferson Lab,}
[{\tt hep-ex/1609.00676}].

\bibitem{Duran:2022xag}
B.~Duran, Z.~E.~Meziani, S.~Joosten, M.~K.~Jones, S.~Prasad, C.~Peng, W.~Armstrong, H.~Atac, E.~Chudakov and H.~Bhatt, \textit{et al.},
\emph{Determining the gluonic gravitational form factors of the proton,}
Nature \textbf{615} (2023) no.7954, 813-816, doi:10.1038/s41586-023-05730-4[{\tt nucl-ex/2207.05212}].

\bibitem{Kharzeev:2021qkd}
D.~E.~Kharzeev, \emph{Mass radius of the proton,} Phys. Rev. D \textbf{104} (2021) no.5, 054015,
doi:10.1103/PhysRevD.104.054015 [{\tt hep-ph/2102.00110}].

\bibitem{Mamo:2019mka}
K.~A.~Mamo and I.~Zahed,
\emph{Diffractive photoproduction of $J/\psi$ and $\Upsilon$ using holographic QCD: gravitational form factors and GPD of gluons in the proton,} Phys. Rev. D \textbf{101}, no.8, 086003 (2020), doi:10.1103/PhysRevD.101.086003, [{\tt hep-ph/1910.04707}].

\bibitem{Guo:2021ibg}
Y.~Guo, X.~Ji and Y.~Liu,
\emph{QCD Analysis of Near-Threshold Photon-Proton Production of Heavy Quarkonium,}
Phys. Rev. D \textbf{103}, no.9, 096010 (2021), doi:10.1103/PhysRevD.103.096010

\bibitem{Hatta:2018ina}
Y.~Hatta and D.~L.~Yang,
\emph{Holographic $J/\psi$ production near-threshold and the proton mass problem,}
Phys. Rev. D \textbf{98} (2018) no.7, 074003, doi:10.1103/PhysRevD.98.074003
[{\tt hep-ph/1808.02163}]

\bibitem{Hatta:2019lxo}
Y.~Hatta, A.~Rajan and D.~L.~Yang,
\emph{Near-threshold $J/\ensuremath{\psi} $and $\Upsilon{}$ photoproduction at JLab and RHIC,} Phys. Rev. D \textbf{100} (2019) no.1, 014032, doi:10.1103/PhysRevD.100.014032
[{\tt hep-ph/1906.00894}].

\bibitem{Sun:2021gmi}
P.~Sun, X.~B.~Tong and F.~Yuan,
\emph{Perturbative QCD analysis of near-threshold heavy quarkonium photoproduction at large momentum transfer,}
Phys. Lett. B \textbf{822} (2021), 136655, doi:10.1016/j.physletb.2021.136655, [{\tt hep-ph/2103.12047}].

\bibitem{Mamo:2022eui}
K.~A.~Mamo and I.~Zahed, \emph{J/\ensuremath{\psi} near-threshold in holographic QCD: A and D gravitational form factors,} Phys. Rev. D \textbf{106}, no.8, 086004 (2022), doi:10.1103/PhysRevD.106.086004 [{\tt hep-ph/2204.08857 hep-ph}].

\bibitem{GlueX:2023pev}
S.~Adhikari \textit{et al.} [GlueX],
\emph{Measurement of the J/$\psi $ photoproduction cross section over the full near-threshold kinematic region,} Phys. Rev. C \textbf{108}, no.2, 025201 (2023), doi:10.1103/PhysRevC.108.025201, [{\tt nucl-ex/2304.03845}].

\bibitem{Guo:2025jiz}
Y.~Guo, F.~Yuan and W.~Zhao,
\emph{Bayesian Inferring Nucleon's Gravitation Form Factors via Near-threshold $J/\psi$ Photoproduction,}
[arXiv:2501.10532 [hep-ph]].

\bibitem{Pefkou:2021fni}
D.~A.~Pefkou, D.~C.~Hackett and P.~E.~Shanahan, \emph{Gluon gravitational structure of hadrons of different spin,} Phys. Rev. D \textbf{105}, no.5, 054509 (2022), doi:10.1103/PhysRevD.105.054509, [{\tt hep-lat/2107.10368}].

\bibitem{Hou:2019efy}
T.~J.~Hou, J.~Gao, T.~J.~Hobbs, K.~Xie, S.~Dulat, M.~Guzzi, J.~Huston, P.~Nadolsky, J.~Pumplin and C.~Schmidt, \textit{et al.}
\emph{New CTEQ global analysis of quantum chromodynamics with high-precision data from the LHC,}
Phys. Rev. D \textbf{103}, no.1, 014013 (2021), doi:10.1103/PhysRevD.103.014013[{\tt hep-ph/1912.10053}]

\bibitem{Guo:2023pqw}
Y.~Guo, X.~Ji, Y.~Liu and J.~Yang, \emph{``Updated analysis of near-threshold heavy quarkonium production for probe of proton\textquoteright{}s gluonic gravitational form factors,} Phys. Rev. D \textbf{108}, no.3, 034003 (2023), doi:10.1103/PhysRevD.108.034003 [{\tt hep-ph/2305.06992}].

\bibitem{Mamo:2021krl}
K.~A.~Mamo and I.~Zahed,
\emph{Nucleon mass radii and distribution: Holographic QCD, Lattice QCD and GlueX data,}
Phys. Rev. D \textbf{103}, no.9, 094010 (2021), doi:10.1103/PhysRevD.103.094010, [{\tt hep-ph/2103.03186}].

\bibitem{JeffersonLabSoLID:2022iod}
J.~Arrington \textit{et al.} [Jefferson Lab SoLID],
\emph{The solenoidal large intensity device (SoLID) for JLab 12 GeV,}
J. Phys. G \textbf{50} (2023) no.11, 110501, doi:10.1088/1361-6471/acda21, [{\tt nucl-ex/2209.13357}].

\bibitem{Gryniuk:2020mlh}
O.~Gryniuk, S.~Joosten, Z.-E.~Meziani and M.~Vanderhaeghen,
\emph{$\Upsilon$ photoproduction on the proton at the Electron-Ion Collider,}
Phys. Rev. D \textbf{102}, no.1, 014016 (2020), doi:10.1103/PhysRevD.102.014016, [{\tt hep-ph/2005.09293}].

\bibitem{Lorce:2018egm}
C.~Lorc\'e, H.~Moutarde and A.~P.~Trawi\'nski,
\emph{Revisiting the mechanical properties of the nucleon,} Eur. Phys. J. C \textbf{79} (2019) no.1, 89
doi:10.1140/epjc/s10052-019-6572-3, [{\tt hep-ph/1810.09837 hep-ph}].

\bibitem{Hackett:2023rif}
D.~C.~Hackett, D.~A.~Pefkou and P.~E.~Shanahan,
\emph{Gravitational Form Factors of the Proton from Lattice QCD,}
Phys. Rev. Lett. \textbf{132} (2024) no.25, 251904, doi:10.1103/PhysRevLett.132.251904, [{\tt hep-lat/2310.08484}].


\bibitem{Hatta:2018sqd}
Y.~Hatta, A.~Rajan and K.~Tanaka,
\emph{Quark and gluon contributions to the QCD trace anomaly,}
JHEP \textbf{12}, 008 (2018), doi:10.1007/JHEP12(2018)008, [{\tt hep-ph/1810.05116}].

\bibitem{Tanaka:2022wzy}
K.~Tanaka, \emph{Twist-four gravitational form factor at NNLO QCD from trace anomaly constraints,}
JHEP \textbf{03}, 013 (2023), doi:10.1007/JHEP03(2023)013 [{\tt hep-ph/2212.09417}].

\bibitem{Prasad:2024}
S. Prasad, Private communication and " Probing the gluonic gravitational form factors of the proton using near-threshold $J/\psi$ photoproduction", Hall C Winter Collaboration meeting, January 18-19, 2024, Newwport News, VA, \url{https://indico.jlab.org/event/758/contributions/13800/}

\end{thebibliography}
\end{document}